\let\accentvec\vec
\documentclass[a4paper]{llncs}

\let\vec\accentvec
\usepackage{MnSymbol} 
\usepackage{graphicx}
\usepackage{times}
\usepackage{listings}
\usepackage{url}
\usepackage{alltt}
\usepackage{tabularx}
\usepackage{verbatim}
\usepackage{enumitem}
\usepackage{footnote} 

\usepackage{url}
\usepackage{microtype}
\newcommand\Small{\fontsize{8.5}{10.5}\selectfont}
\newcommand*\LSTfont{\Small\ttfamily} 

\begin{document}

\title{Embedding agents in business applications using enterprise integration patterns}

\author{Stephen Cranefield \and Surangika Ranathunga}
\institute{Department of Information Science, University of Otago, Dunedin, New Zealand\\
           \{scranefield,surangika\}@infoscience.otago.ac.nz}

\maketitle

\begin{abstract}
This paper addresses the issue of integrating agents with a variety of external resources and services, as found in enterprise computing environments. We propose an approach for interfacing agents and existing message routing and mediation engines based on the \emph{endpoint} concept from the enterprise integration patterns of Hohpe and Woolf. A design for agent endpoints is presented, and an architecture for connecting the Jason agent platform to the Apache Camel enterprise integration framework using this type of endpoint is described. The approach is illustrated by means of a business process use case, and a number of Camel routes are presented. These demonstrate the benefits of interfacing agents to external services via a specialised message routing tool that supports enterprise integration patterns.
\end{abstract}

\section{Introduction}

Much of the research in multi-agent systems (MAS) is based on a conceptual model in which the only entities are agents and an abstracted external environment. This is in contrast to modern enterprise computing environments, which comprise a diverse range of middleware and server technologies.


The current solutions for integrating agents with external computing infrastructure are: (a) to access these resources and services directly from agent code (if using a conventional programming language), (b) to implement user-defined agent actions %
or an environment model %
to encapsulate these interactions, (c) to provide custom support in an agent platform for specific types of external service, or (d) to provide a generic interface for calling external resources and services, either using a platform-specific API \cite{IMPACT_JIIS_2000} or by encapsulating them as agents~\cite{GeneserethKetchpel}, artifacts~\cite{JAAMAS_artifacts} or active components~\cite{ActiveComponentsMATES10}. However, none of these approaches are a good solution when agents need to be integrated with a range of technologies. They either require agent developers to learn a variety of APIs, or they assume that agent platform developers or their users will provide wrapper templates for a significant number of commonly used technologies.

\begin{figure}[!tb]
\begin{minipage*}{\textwidth}
\centerline{
\includegraphics[width=0.8\textwidth]{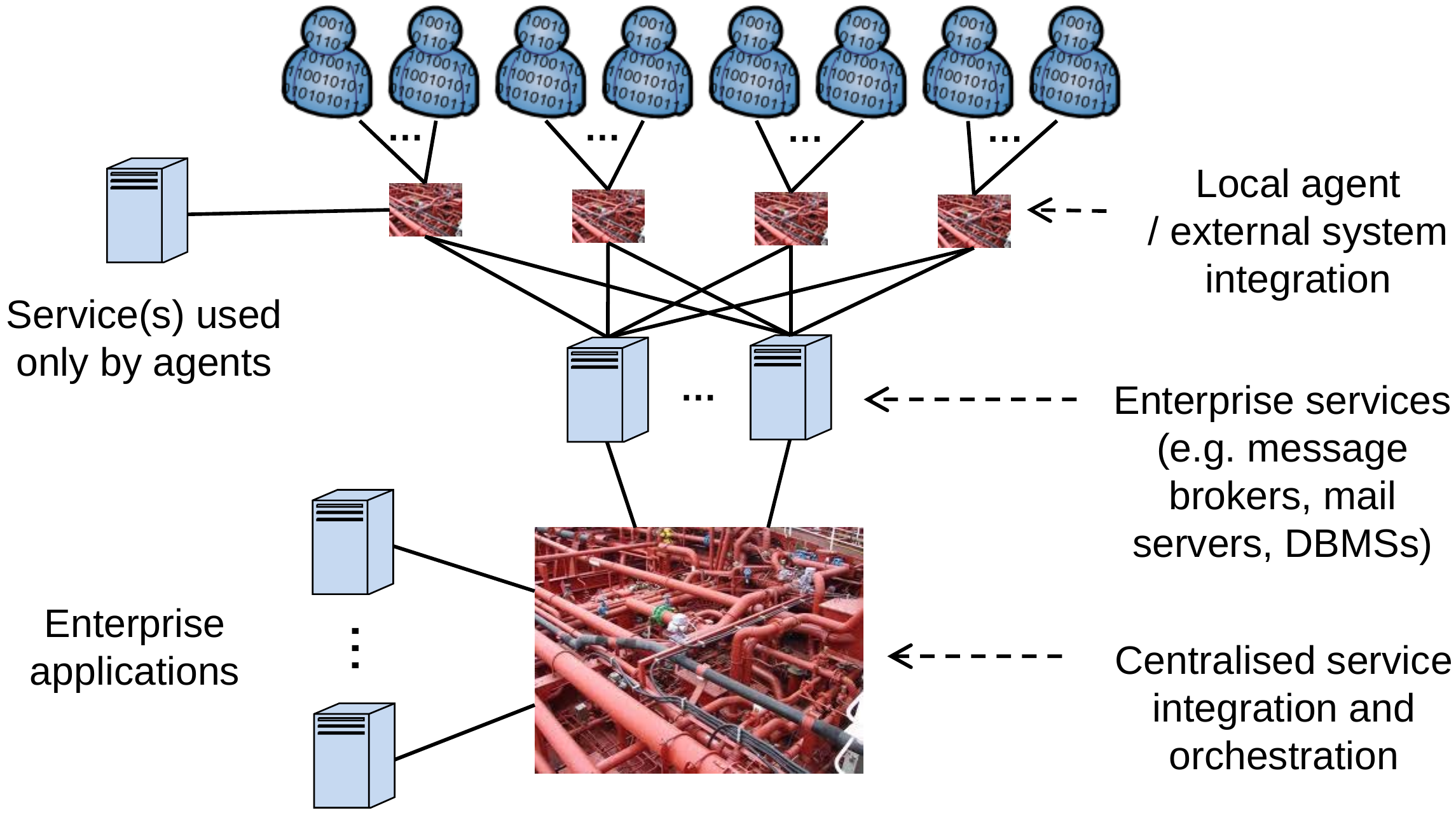}}
\caption{The proposed MAS integration model\protect\footnotemark}
\label{fig:my_model}
\end{minipage*}
\end{figure}

This paper proposes an alternative approach: the use of a direct bridge between agents and the mainstream industry technology for enterprise application integration: message routing and mediation engines, and in particular, those that support the enterprise integration patterns (EIP) of Hohpe and Woolf~\cite{HohpeWoolf2004}. Our integration approach is illustrated in Figure~\ref{fig:my_model}. In this figure, each ``pipes'' graphic represents a messaging-based service coordination tool, such as an enterprise service bus~\cite{ESBbook}. The larger one represents an organisation's existing message-based infrastructure for managing business processes by coordinating information passing between applications and services. We propose that agents can be embedded into this infrastructure by integrating them with their own local message-routing and mediation engines, such as the lightweight Java-based Apache Camel enterprise integration framework~\cite{CamelBook}. This integration is based on the EIP notion of an \emph{endpoint}, and we present the design of endpoints that can translate agent requests (encoded as agent communication language messages or action executions) to EIP messages, and from EIP messages to agent messages and percepts.

\footnotetext{Pipes photo by Herv\'e Cozanet, source: \protect\url{http://commons.wikimedia.org/wiki/File:Piping_system_on_a_chemical_tanker.jpg} (CC BY-SA 3.0)}

We describe an implemented architecture for connecting the Jason agent platform~\cite{JasonBook} to Camel using these ``agent endpoints''. The approach is illustrated by means of a business process use case requiring the integration of Jason agents with a database management system, a mail server, a message broker and the Apache ZooKeeper coordination server. A number of Camel routes handling aspects of this use case are presented to demonstrate the benefits of interfacing agents to external services via a specialised message routing tool that supports enterprise integration patterns.

\section{Enterprise Integration Patterns}

Enterprise computing environments typically comprise hundreds and possibly thousands of applications \cite{HohpeWoolf2004}. These may use a variety of communication protocols and interface technologies due to departmental autonomy (e.g.\ to acquire ``best of breed'' applications for specific business problems), incremental and opportunistic growth, mergers, etc. To preserve loose coupling between the diverse applications involved in the automation of business processes, and thus facilitate maintenance and extensibility, the use of middleware products based on asynchronous message-passing has emerged as the mainstream approach for \emph{enterprise application integration}. In this approach, applications interact by sending and receiving structured messages to and from named queues or publish-subscribe `topics' managed by (possibly federated) \emph{message brokers}. Message routing and transformation rules can be executed by the message broker or by specialised message routing and mediation engines, thus providing a single location for the specification of business processes. The concept of the \emph{enterprise service bus} extends this idea further by integrating message brokers with middleware for deploying and interacting with various type of service, such as web services~\cite{ESBbook}.

Hohpe and Woolf \cite{HohpeWoolf2004} have identified 65 ``enterprise integration patterns'' (EIPs) for solving basic problems that commonly arise in messaging-based enterprise application integration, such as the \emph{scatter-gather} pattern: ``How do you maintain the overall message flow when a message needs to be sent to multiple recipients, each of which may send a reply?'' A number of middleware tools have direct support for these patterns, including Apache Camel.

\section{Apache Camel}

Camel is an open source Java framework for executing message routing and mediation rules that are defined using domain-specific languages (DSLs) based on Java and Scala, or by using XML configuration files. In the work reported in this paper we have used the Java DSL.

Camel is based on the EIP concepts of \emph{routes} and \emph{endpoints}. A Camel application comprises a set of \emph{route definitions}. Each route receives messages from a \emph{consumer endpoint}, and performs a sequence of processing steps on each message, such as filtering and transforming messages, before sending the processed messages to one or more \emph{producer endpoints}. Endpoints can be ``direct'' links to other routes in the application (i.e.~messages leaving one route may flow directly into another route) or they may represent connections to external resources and services. For example, a \emph{mail} endpoint may be used as a consumer to receive messages representing unread mail in a specified account on a mail server, or as a producer that sends mail to a server. Camel has more than 130 different \emph{components} defined to provide a variety of endpoint types. These enable sending and/or receiving messages to and from external resources such as files, databases, message brokers, generic web services, specific Amazon and Google services, RSS and Atom feeds, and Twitter. To enable this diversity of endpoint types, Camel's concept of a message is very general: a message has headers (a map from names to Java objects), a body (which can be any Java object) and optional attachments. 
%
%

The code below defines two simple Camel routes. These use the agent component described in this paper to enable ``local'' agents (those running within the same process as the Camel routes) to communicate with remote agents via a message broker.
\label{sec:jms_routes}

\begin{alltt}
\textbf{from}("agent:message")
\textbf{.setHeader}("CamelJmsDestinationName",
   \textbf{simple}("\${headers.receiver.split(\textbackslash"\textunderscore\textunderscore\textbackslash")[0]}"))
\textbf{.to}("jms:dummy")

\textbf{from}("jms:"+containerId)\textbf{.to}("agent:message");
\end{alltt}

These routes are defined using Camel's Java DSL. This is a Java API for constructing routes via a sequence of method calls. The \verb|from| method creates a consumer endpoint and the \verb|to| method creates a producer endpoint. Endpoints are specified using uniform resource identifiers (URIs), with the first part of the URI (the scheme) identifying the type of the endpoint. Other parts of the URI provide additional details, and the various endpoint types provided by Camel make use of URI parameters to provide configuration details for the instantiation of the endpoint. The routes shown above use two types of endpoint: the agent endpoint described in this paper, and Camel's JMS endpoint for sending and receiving messages from a message broker using the Java Message Service.

The first route definition above creates an endpoint that receives all messages sent by local Jason agents. For each Jason message received, this endpoint copies the message content into the body of a new Camel message, and records the other message details using \verb|sender|, \verb|receiver| and \verb|illoc_force| headers (these correspond directly to Jason message properties).

\begin{sloppypar}
The routes are run within a Camel \emph{context} object. Our architecture allows multiple distributed Camel contexts, each with their own set of local agents running within an \emph{agent container}, so all agents are created with names of the form \verb|containerId__localName|. The second and third lines of the first route above use Camel's ``Simple'' expression language to extract the first part of the name, which identifies the agent container that the message recipient is attached to, and stores this as the value of a specific header predefined by the JMS component. When the message is processed by the JMS producer endpoint, this header is used to override the queue or topic name that appears as a mandatory component of a JMS endpoint URI (hence the ``dummy'' message queue name at the end of the first route above). This illustrates two aspects of the use of message headers in Camel: they are commonly used within routes to store information needed later in the route, and they can affect the handling of messages by endpoints.
\end{sloppypar}

The second route definition above creates a JMS endpoint that receives messages from a message broker (the address of the broker is provided to Camel's JMS component on initialisation). The endpoint listens to a specific queue, which is named after the unique identifier for the local agent container (note that there may be agent containers associated with other Camel contexts running elsewhere on the network or in other processes). The JMS consumer endpoint copies the body and the message headers from the received JMS messages to create Camel message objects. The route specifies that these messages flow from the JMS consumer endpoint directly to a Jason producer endpoint. This endpoint generates Jason messages corresponding to the Camel messages and delivers them to the appropriate agents. The Jason producer endpoint does the reverse of the Camel to Jason message mapping described above.

Camel supports a number of \emph{message exchange patterns} (MEPs), with the most commonly used being \verb|InOnly| and \verb|InOut|. The pattern to use for handling a message arriving at a consumer endpoint is set by that endpoint, possibly based on information in the message (such as a \verb|JMSReplyTo| header on incoming JMS messages). The MEP can also be manually set by a route using methods of the Java DSL. If a message reaches the end of a route with the \verb|InOut| MEP, it is returned to the consumer endpoint. If that endpoint supports it, that message will be treated as the reply to the initial request. Thus Camel can be used to implement both synchronous and asynchronous processing of messages.

Note that Camel routes can be significantly more complicated than those shown above, as later examples in this paper will demonstrate. In particular, the Java DSL includes methods for conditional branching, exception-handling and for starting, stopping, suspending and resuming routes. In addition, an important feature of Camel is the provision of methods that can be used singly or in combination to implement enterprise integration patterns such as splitting and aggregating messages, or to ``enrich'' messages with content obtained by making synchronous calls to other endpoints.

\section{A Jason/Camel bridge}
\label{sec:integration}

\begin{figure}[!tb]
\centering\includegraphics[width=0.8\textwidth]{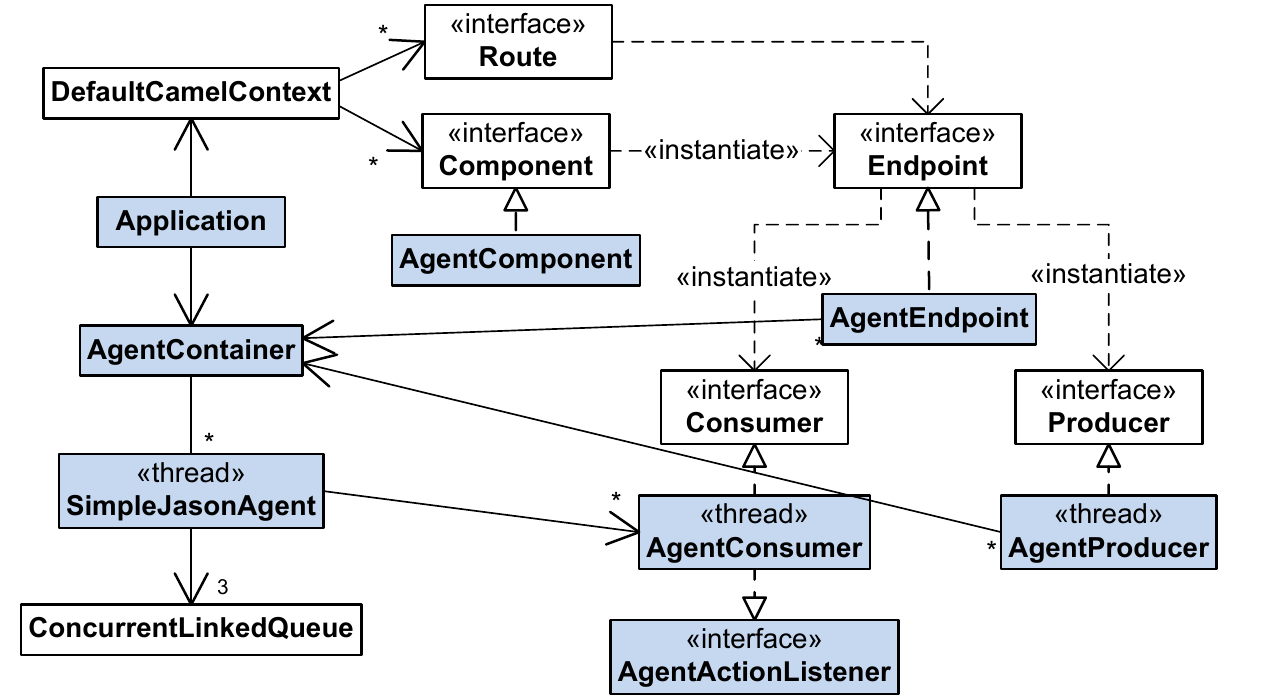}
\caption{The architecture of our Jason/Camel bridge}
\label{fig:architecture}
\end{figure}

In this section we briefly describe the architecture of our Jason/Camel bridge and discuss how we map between the conceptual models of Jason and Camel. In particular, we describe the design and interpretation of agent endpoints.

\subsection{Application architecture}
\label{sec:architecture}
\begin{sloppypar}
Our Jason/Camel bridge\footnote{\url{http://github.com/scranefield/camel-agent}} consists of an ``agent component'' for Camel and an application template that integrates the Jason BDI interpreter with a Camel context. The agent component for Camel is a factory for creating agent consumer and producer endpoints. Its implementation consists of the component class and classes that are instantiated to create producer and consumer endpoints for Jason.
\end{sloppypar}

The architecture of the bridge is shown in Figure~\ref{fig:architecture} using UML\footnote{Classes and interfaces developed or adapted by us are shaded in the figure.}. A Camel application initialises any required components, creates a CamelContext object, passes it a \verb|Route|-\verb|Builder| object with a method that defines the routes, and then starts the context. Our integration architecture extends this by adding to the application an agent container. On initialisation, this container locates all Jason agent source (.asl) files in a given directory\footnote{This simple approach will be replaced in the future by the use of OSGi ``bundles'' to package and deploy Camel contexts together with their associated agents.} and, for each agent, instantiates our extension of the \verb|SimpleJasonAgent| class\footnote{\url{http://jason.sourceforge.net/faq/faq.html#SECTION00057000000000000000}}. This class allows the Jason BDI interpreter to be used without any of the existing Jason ``infrastructures'' for agent communication. It is responsible for providing the BDI interpreter with methods to call to get percepts, to perform actions, and to send and check for messages. We chose this as the most lightweight approach for embedding Jason agents into business processes via Camel%
%

Our \verb|SimpleJasonAgent| class maintains concurrently accessible queues for percepts of two types (\emph{transient} and \emph{persistent}) and for incoming messages. Messages on these queues are read (and consumed in the case of transient percepts) when the BDI interpreter calls the class's methods for getting percepts and messages. Note that each agent and endpoint runs in a separate thread. The agent container writes messages and percepts to the queues for the relevant agents after receiving them from \verb|agent:message| and \verb|agent:percept| endpoints that appear in Camel routes. An endpoint for producing percepts chooses whether percepts are transient or persistent based on the endpoint URI parameters and/or the headers of the Camel message being processed. Transient percepts are cleared after an agent has perceived them, whereas persistent ones will repeatedly perceived (but may be overwritten by other percepts with the same functor---see the discussion of the \verb|updateMode| URI parameter and message header in Section~\ref{sec:endpoint-design}).

On construction, each agent is passed a list of agent consumer endpoints, and these are used to deliver messages and actions---the endpoints are responsible for selecting which of these match their configuration parameters. Camel messages generated by the consumer endpoints are processed using the \verb|InOnly| message exchange pattern, unless specified otherwise by a route or an endpoint URI.

Inter-agent messaging via a message broker, as implemented by the routes shown above in Section~\ref{sec:jms_routes}, requires the existence of a separate message queue for each agent container. To enable this functionality, our application class has a optional configuration parameter specifying that an Apache ZooKeeper\footnote{\url{http://zookeeper.apache.org/}} server should be used to dynamically obtain a unique identifier for the container.

\label{sec:zookeeper}
A ZooKeeper server maintains a set of named nodes, arranged in a tree structure, to which system configuration information can be read and written by clients. The nodes are kept in memory to enable high performance, but transaction logs and persistent snapshots are also used to provide reliability. The data can be replicated across a cluster of ZooKeeper servers. Nodes can be persistent or \emph{ephemeral}---a node of the latter type is automatically deleted if the client session that created it is no longer maintaining a ``heartbeat''. Nodes can also be \emph{sequential}. These have a unique number appended to the specified node name, based on a counter associated with the parent node. A client can place a \emph{watch} for changes to the data recorded in a node, the existence of a node, or the set of children of a node. Together, these features can be used to implement a range of distributed coordination mechanisms, such as distributed queues, barriers and locks, maintaining lists of active group members, and electing group leaders.

Our application class obtains the agent container identifier by requesting the creation of an ephemeral sequence node with the path \verb|containers/container| and receives in response the name of the created node with a sequence number appended. 

ZooKeeper servers can also be accessed from within Camel routes, via ZooKeeper endpoints. A use case for this functionality is illustrated in our MAS application scenario in Section~\ref{sec:case_study}.

Another option provided by our bridge is to \emph{directly} deliver messages between agents that are in the same context, if preferred, rather than sending these to the Camel context for routing via JMS or any other means specified by the provided routes.

\begin{table}[t]
\begin{tabularx}{\textwidth}{|l|>{\raggedright\arraybackslash}X|>{\raggedright\arraybackslash}X|>{\raggedright\arraybackslash}X|}
\multicolumn{4}{c}{\textbf{Consumer endpoints}}\\[4pt]\hline
\textbf{Endpoint type} & \textbf{Optional parameters} & \textbf{Camel headers set} & \textbf{Camel body contains}\\\hline
\texttt{agent:message} & 
\texttt{illoc\textunderscore{}force}, \texttt{sender}, \texttt{receiver}, \texttt{annotations}, \texttt{match}, \texttt{replace} & 
\texttt{illoc\textunderscore{}force}, \texttt{sender}, \texttt{receiver}, \texttt{annotations}, \texttt{msg\textunderscore{}id} & 
The message content\newline (as a string)\\\hline
\texttt{agent:action} & \texttt{actor}, \texttt{annotations}, \texttt{match}, \texttt{replace}, \texttt{resultHeaderMap} & 
\texttt{actor}, \texttt{annotations}, \texttt{actionName}, \texttt{params} & 
The action term\newline (as a string)\\\hline
\multicolumn{4}{c}{}\\
\multicolumn{4}{c}{\textbf{Producer endpoints}}\\[4pt]\hline
\textbf{Endpoint type} & \textbf{Optional parameters} & \textbf{Camel headers used} & \textbf{Camel body expected to be} \\\hline
\texttt{agent:message} & \texttt{illoc\textunderscore{}force}, \texttt{sender}, \texttt{receiver}, \texttt{annotations} & 
\texttt{illoc\textunderscore{}force}, \texttt{sender}, \texttt{receiver}, \texttt{annotations} &
The message content\newline (as a string)\\\hline
\texttt{agent:percept} & \texttt{receiver}, \texttt{annotations}, \texttt{persistent}, \texttt{updateMode} &
\texttt{receiver}, \texttt{annotations}, \texttt{persistent}, \texttt{updateMode} &
The percept (as a string)\\\hline
\end{tabularx}
\caption{Agent endpoint types}
\label{table:jason-endpoint-types}
\end{table}

\subsection{Agent endpoint design}
\label{sec:endpoint-design}
We support two types of Jason \emph{consumer} endpoints to handle local agent messages and actions delivered to them from our Jason/Camel bridge. Endpoints of these types generate Camel messages that correspond (respectively) to \emph{messages} sent by the local agents and to \emph{actions} executed by them. The details of the Jason messages and actions are encoded in the headers and body of the Camel message, as shown in Table ~\ref{table:jason-endpoint-types}. For example, the content of a Jason message is placed in the body of the Camel message, and the \verb|illoc_force| (illocutionary force), \verb|sender|, \verb|receiver|, \verb|msg_id| and \verb|annotations| properties of the Jason message are stored on the Camel message using headers with these names. 

\begin{sloppypar}
A route definition creates these types of endpoints by calling the \verb|from| method with an argument that is string of the form \verb|"agent:message?options"| or \verb|"agent:action?options"|. The options are specified using the standard URI query parameter syntax \verb|?opt1=v1&opt2=v2...|. Camel messages are only generated by these endpoints if the selection criteria specified by the optional parameters are satisfied. The parameters recognised by these endpoint types are shown in Table~\ref{table:jason-endpoint-types} and explained below.
\end{sloppypar}

We also support two types of Jason \emph{producer} endpoints, which generate \emph{messages} and \emph{percepts}, respectively, for the local agents. These messages and percepts are created from Camel messages that reach the endpoints via Camel routes, and their content is taken from the body and headers of those Camel messages and the endpoint URI parameters. As shown in Table~\ref{table:jason-endpoint-types}, the URI parameters supported for the producer endpoints are mirrored by the headers that the endpoints check. This is because these message headers can be used to override the URI parameters when converting a Camel message to a Jason message or percept. This allows Camel routes to dynamically control the delivery and construction of Jason messages and percepts.

The URI endpoint parameters and Camel message headers are used as agent message and action selectors (for consumer endpoints) or to specify generated percepts or agent messages (for producer endpoints). Below, we provide some additional details for some of the parameter and header options.

\begin{description}[leftmargin=0pt, style=sameline]
\item[\texttt{receiver:}] We interpret the value ``\verb|all|'' for this URI parameter and message header as meaning that only broadcast messages should be selected by a message consumer endpoint or that the message should be sent to all local agents from a message or percept producer endpoint. This is the default value for a producer. No agent can have this name because the agent container identifier is prepended to the names of all agents on creation. The \verb|receiver| value can also be a comma-separated list of recipients when provided to a message or percept producer endpoint.

\item[\texttt{annotations:}] Jason supports the attachment of a list of \emph{annotation} terms to a literal. An \verb|annotations| URI parameter or header can be specified for controlling the selection of messages or actions by a consumer endpoint or to trigger the generation of annotations by a producer endpoint. The values are specified as a comma-separated list of literal strings (for the parameter) or as a Java list of strings (when using a header).

\item[\texttt{match} and \texttt{replace}:] These are used on consumer endpoints. A \verb|match| parameter specifies a regular expression, and a Camel message is only generated if this matches the incoming message or action (in string format). The Java regular expression syntax is used, and pairs of parentheses may be used to specify `groups' in the pattern. The values corresponding to these groups in the matched string are recorded and used when processing a \verb|replace| parameter (if present). A \verb|replace| parameter specifies a string to be used as the body of the generated Camel message. This can contain group variables (in the form \verb|$|$n$), and these are replaced with the values that were recorded during matching.

\item[\texttt{resultHeaderMap:}] An action consumer endpoint supports both synchronous and asynchronous actions. An asynchronous action corresponds to a Jason \emph{external} action (which cannot contain variables), and the endpoint always returns the result \verb|true| to the agent that performed the action. In order to handle actions that map to routes with an \verb|InOut| message exchange pattern, we implemented a Java class that provides a new Jason internal action \verb|jasoncamel.syncInOut|-\verb|Exchange|. This is used to send terms that represent actions with unbound variables to Jason action consumer endpoints. Once the route (which is run with an \verb|InOut| message exchange pattern) is completed, the endpoint unifies the variables in the action term with the resulting Camel message body. An endpoint processing this type of action must have a \verb|resultHeaderMap| endpoint parameter. Its value should be a comma-separated list of \emph{header-name}\verb|:|\emph{argument-index} pairs. When a Camel message completes the route, for each of these pairs the value of the header with name \emph{header-name} header is unified with the action term's argument at index \emph{argument-index}.

\item[\texttt{persistent} and \texttt{updateMode}:] As described in Section~\ref{sec:architecture}, percepts delivered to agents by a percepts producer endpoint can be transient or persistent. The choice is controlled by the \verb|updateMode| URI parameter or a message header with that name. Persistent percepts with the same functor and arity but different argument values can either \emph{accumulate} in an agent's persistent percepts list, or each new percept of that form can \emph{replace} previous ones. The latter case is useful for percepts that represent the state of an external resource. A value of ``\verb|-+|'' for an \verb|updateMode| URI parameter or a Camel message header with that name can be used to specify the percept replacement behaviour. This can also apply to transient beliefs to prevent multiple percepts with the same functor and arity being queued up between consecutive perceptions by an agent.
\end{description}

\begin{figure}[!tb]
\centerline{\includegraphics[width=0.8\textwidth]{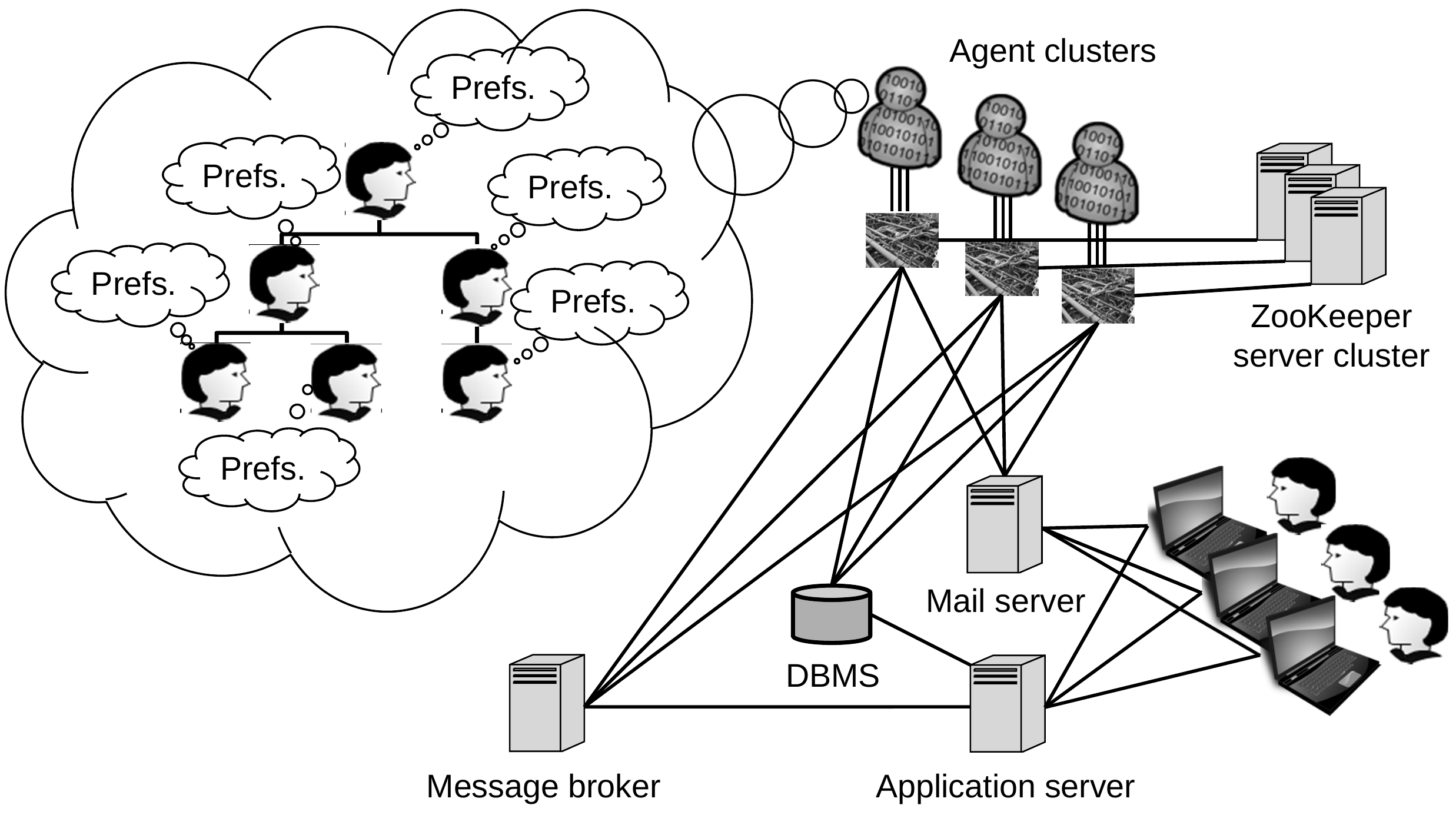}}
\caption{Architecture of our use case}
\label{fig:case_study}
\end{figure}

\section{A business process use case}
\label{sec:case_study}

In this section we illustrate the use of the Jason/Camel bridge by describing a hypothetical business process in which agents could play a valuable role. This use case addresses the problem of achieving more targeted information flow within an organisation and reducing the overuse of the CC header in email messages. Our solution, shown in Figure~\ref{fig:case_study}, assumes the existence of a specific ``to.share'' email account. Users with information they think may be of interest to others can mail it to this account. Agents monitor this account and evaluate the relevance of each new message to other users, with each agent responsible for considering the interests and needs of a specific subset of users. The sets of users assigned to the agents form a partition of the complete user base. The agents base their decision on knowledge of the roles of users and the organisational structure (stored in a database), as well as specific plans that may optionally be provided by users to encode their goals for receiving information. We assume that these plans are created using a graphical web interface that provides a user-friendly abstraction layer on top of Jason's plan syntax. When agents determine which users might be interested in an email message, they deliver the message to those users' mail accounts via SMTP.

\lstset{%
basicstyle=\LSTfont, %
morecomment=[l][\itshape]{//\ }, %
keywordstyle=\bfseries, %
breaklines=true, %
showtabs=false, %
upquote=true, %
showstringspaces=false, %
prebreak=\raisebox{0ex}[0ex][0ex]{\ensuremath{\rhookswarrow}}, %
keywordsprefix={from}, %
morekeywords={to,header,body,constant,setBody,setHeader,simple,groovy,idempotentConsumer,eager,split,process,aggregate,completionSize,completionTimeout}, %
escapechar=|,%
numberstyle=\tiny, stepnumber=1, numbersep=7pt}

\begin{lstlisting}[float=bt,frame=lines,caption=Camel route for implementing an action as a database query,label=listing:account-query]
from("|\underline{\LSTfont agent:action}|?exchangePattern=InOut" +
                 "&actionName=get_email_accounts" +
                 "&resultHeaderMap=result:1"        )
  .setBody(constant("select email |{\LSTfont from}| users"))
  .to("jdbc:dataSource")
  .setHeader("result").groovy(
     "exchange.in.|{\LSTfont body}|.collect{'\"'+it['email']+'\"'}")
\end{lstlisting}

Our system design for implementing this business process involves coordinated use of Jason agents, a mail server, a database management system, a message broker and ZooKeeper, with the coordination performed by Camel routes. The key routes are as follows\footnote{Note that the two example routes presented earlier in Section~\ref{sec:jms_routes} are not used because agents in this application do not send messages to each other, but rather interact with external services via actions and percepts.}.

\begin{enumerate}
\item
On start-up, each agent performs an action \verb|get_email_| \verb|accounts(Accounts)| that is mapped to a database query by a Camel route. The route sets a message header to hold the list of accounts, and the agent consumer endpoint instantiates the argument \verb|Accounts| with this value. After this action succeeds, the agent records this account list in a belief. This route is shown in Listing~\ref{listing:account-query}.
\item
On start-up, each agent also performs a \verb|register| action. A route maps this to the creation of an ephemeral sequential node in ZooKeeper (under the node \verb|/agents|).
\item
A route is watching the children of the ZooKeeper node \verb|/agents|. Whenever there is a change (due to Camel contexts and their associated agent containers starting and stopping), the route sends an updated list of active agents to its local agents as a persistent percept in \verb|-+| update mode. This, and the route described in the previous paragraph, are shown in Listing~\ref{listing:zookeeper-routes}.
\item
Whenever a new email account is created or deleted by the administrator, the database is updated, and in addition a notification of the change is sent to a specific publish-subscribe topic on a message broker (a topic is needed rather than a queue to allow \emph{all} running Camel contexts to receive the message). A route monitors this topic and sends any received messages as transient percepts to all local agents.
\item
Similar routes are provided for agents to obtain from the database information about users' roles and their positions in the organisation structure, as well as a set of default email-forwarding plans. This is done by database queries from routes that are triggered by agent actions. Notifications of changes to this information are sent by an administrative tool (or a database trigger) to a message broker topic. A route monitoring that topic generates updates to the corresponding persistent agent percepts, which then cause agents to call the actions to load this information again.
\item
Similar routes are also provided for agents to retrieve from a database, for a specified set of users, their information relevance assessment plans, and to receive notifications of changes to these plans via a message broker topic.
\item
The agents have plans that react to changes in their beliefs about the currently active agents and the list of email accounts. When a change occurs, they each run an algorithm (common to all agents) to divide the list of email accounts amongst them, based on their own position in the list of agents. They maintain a belief recording the accounts they are responsible for.
\item
A set of routes polls the ``to.share'' email account for new mail using a mail consumer endpoint, sends a message to all local agents asking them to evaluate which of their allocated email accounts the message is relevant for, aggregates the reply with the email message, and forwards it to the nominated users, via a mail producer endpoint. These routes are shown in Listing~\ref{listing:mail-routes}.
\item
When the list of accounts that an agent is responsible for changes, or it is notified of changes to the plans for any of the accounts it handles, it must re-fetch plans for the relevant agents. The routes handling these notifications suspend the mail-polling route and start another route that uses a timer endpoint to resume the mail-polling route after a fixed amount of time%
. This gives agents time to fetch any required new plans from the database.
\end{enumerate}

\begin{lstlisting}[float=*tbp,numbers=left,frame=lines,caption=Camel routes for tracking active agents via ZooKeeper,label=listing:zookeeper-routes]
// Implement registration by creating a new ZooKeeper sequence
// node with the agent name as its content
from("|\underline{\LSTfont agent:action}|?actionName=register")
  // Process only one register action |{\LSTfont\itshape from}| each agent
  .idempotentConsumer(
     header("actor"),
     MemoryIdempotentRepository.memoryIdempotentRepository(100)
   ).eager(true)
  .setBody(header("actor")) // Put actor name in message |{\LSTfont\itshape body}|
  .to("zookeeper://" + zkserver + "/agents/agent" +
        "?create=true&createMode=EPHEMERAL_SEQUENTIAL");

// Watch agents node in ZooKeeper for changes |{\LSTfont\itshape to}| list of children
from("zookeeper://" + zkserver + "/agents" +
       "?listChildren=true&repeat=true")
  .setHeader("numChildren", |{\LSTfont simple}|("${|{\LSTfont body}|.size}"))
  .split(body()) // Split agent node list into separate messages
  .process(new Processor() {
     public void |{\LSTfont process}|(Exchange exchange) throws Exception {
       // Map the ZooKeeper node name for an agent |{\LSTfont\itshape to}| the agent
       // name by getting the content of the ZooKeeper node
       ConsumerTemplate consumer = camel.createConsumerTemplate();
       String agentName =
         consumer.receiveBody("zookeeper://"+zkserver+"/agents/"
                                + exchange.getIn().getBody(),
                              String.class);
       exchange.getIn().|{\LSTfont setBody}|(agentName);
     }})
  // Aggregate mapped names into a single message containing a
  // list of names. All messages will have the same headers - any
  // will do as the message correlation id
  .aggregate(header("numChildren"),
             new ArrayListAggregationStrategy()
   ).completionSize(header("numChildren"))
  .setBody(simple("agents(${bodyAs(String)})"))
  .to("|\underline{\LSTfont agent:percept}|?persistent=true&updateMode=-+");
\end{lstlisting}

\begin{lstlisting}[float=*tbp,numbers=left,frame=lines,caption=Camel routes for forwarding email based on agent recommendations,label=listing:mail-routes]
// Poll for email messages
from("imaps://mail.bigcorp.com?username=|{\LSTfont to}|.share"
      + "&password="+mailPassword+"&delete=true&copyTo=processed")
  .setHeader("id", simple("\"${id}\""))
  .to("seda:forward-message", "direct:ask-agents");

// Request agents |{\LSTfont\itshape to}| evaluate message on behalf of their
// allocated users
from("direct:ask-agents")
  .setBody(
     simple("check_relevance(" +
              "${|{\LSTfont header}|.id}, \"${|{\LSTfont header}|.|{\LSTfont from}|}\", " +
              "\"${|{\LSTfont header}|.subject}\", \"${bodyAs(String)}\")"))
  .setHeader("receiver", constant("all"))
  .setHeader("sender", constant("router"))
  .to("|\underline{\LSTfont agent:message}|?illoc_force=achieve");

// Receive responses |{\LSTfont\itshape from}| agents and |{\LSTfont\itshape aggregate}| them |{\LSTfont\itshape to}| get a
// single lists of relevant users
from("|\underline{\LSTfont agent:message}|?illoc_force=tell" +
                  "&receiver=router" +
                  "&match=relevant\\((.*),(.*)\\)" +
                  "&replace=$1:$2")
  .setHeader("id", simple("${|{\LSTfont body}|.|{\LSTfont split}|(\":\")[0]}"))
  .setBody(simple("${|{\LSTfont body}|.|{\LSTfont split}|(\":\")[2]}"))
  .aggregate(header("id"),
             new SetUnionAggregationStrategy()
   ).completionTimeout(2000)
  .setHeader("|{\LSTfont to}|", simple("${bodyAs(String)}"))
  .to("seda:forward-message");

// Aggregate original mail message with message summarising
// interested users in "|{\LSTfont\itshape to}|" |{\LSTfont\itshape header}|, and send it
from("seda:forward-message")
  .aggregate(header("id"),
             new CombineBodyAndHeaderAggregationStrategy("|{\LSTfont to}|")
   ).completionSize(2)
  .setHeader("from", constant("|{\LSTfont to}|.share@bigcorp.com"))                          
  .to("smtp://|{\LSTfont to}|.share@mail.bigcorp.com?password="+mailPassword);
\end{lstlisting}

Listings~\ref{listing:account-query}, \ref{listing:zookeeper-routes} and \ref{listing:mail-routes} show the routes we have implemented and tested for three aspects of the system's functionality. We underline the beginnings of the agent endpoint URIs to highlight where the integration with agents occurs. Listing~\ref{listing:account-query} shows how the execution of an agent action literal with a free variable can be implemented by a Camel route with the \verb|InOut| message exchange pattern (note the use of the standard Camel URI parameter \verb|exchangePattern|). The route sends an SQL query to a pre-configured database connection, the returned result is converted to an AgentSpeak list of strings using a Groovy expression, and then the \verb|result| header is used to store the result. The consumer endpoint URI has a \verb|resultHeaderMap| parameter specifying that the endpoint should unify the value of the \verb|result| header with the first argument of the action literal.

Listing \ref{listing:zookeeper-routes} illustrates how Camel provides a convenient way to use ZooKeeper to monitor the active members of a distributed group of agents, and to map this information to agent percepts. The first route (lines 3--11) implements the agent \verb|register| action by creating an ephemeral sequential node (see Section~\ref{sec:zookeeper}) in a ZooKeeper server to represent the agent, and storing its name in that node. Camel's support for the \emph{idempotent receiver} enterprise integration pattern provides a simple way to filter out duplicate registration requests from agents. The second route (lines 14--36) is triggered by changes to the set of ZooKeeper sequence nodes representing agents. On each change, it receives a message listing the current sequence nodes. The \emph{splitter} pattern is used (line 17) to obtain a separate message for each node, and each of these triggers a query to ZooKeeper to get the agent name stored at that node (lines 22--27). Finally (lines 32--34), the \emph{aggregator} pattern is used to combine the names into a list stored in the body of a single message, and that is sent to the local agents as the argument of a percept (lines 35--36).

In the first route in Listing \ref{listing:mail-routes}, the \emph{to.share} mail account is polled for new mail (lines 2--3). A Camel message representing each new mail message is generated and the Camel message exchange identifier is written to a message header for latter use in correlating the agent responses with this Camel message (line 4). The message is then forwarded to two other routes (line 5). One is started asynchronously (via a ``seda'' endpoint, which queues incoming messages) and the other synchronously (via a ``direct'' endpoint). The second route (lines 9--16) sends an \emph{achieve} request to the local agents, asking them to consider whether the mail is relevant to any of their allocated users. The third route (lines 20--28) handles messages sent by agents in response to this goal, which contain lists of potentially interested users. The \emph{aggregator} pattern (lines 24--26) is used to produce, for each email message, a single message containing a combined list of users to forward it to. This is sent to the final route (lines 32--37), which also has (in a queue) the Camel message containing the email message that is waiting to be forwarded. This route uses the \emph{aggregator} pattern again to combine the email message and the list of users to send the message to (stored in the \verb|to| header). Finally, an SMTP endpoint is used to send the mail to these users.

The routes discussed in this paper have been tested using Jason stubs and the necessary external services, but the full Jason code for this scenario has not yet been developed and is not the focus of this paper. However, because the coordination logic is factored out and encoded in the Camel routes, the agent code required will be much simpler than would otherwise be needed without the use of our Jason/Camel bridge. Most of the agent behaviour is to react to percepts sent from Camel by performing actions (e.g. to fetch an updated list of email accounts), and to use Jason's \verb|.add_plan| and \verb|.remove_plan| internal actions to update the plans used to evaluate the relevance of email messages to users. In response to the goal to evaluate a message, the agent must call the user plans, collect the users for whom these plans succeed, and send these in a message to Camel. The agents must also recompute the allocation of users to agents whenever the set of agents changes or new users are added, which they detect via `new belief' events.

\section{Related Work}

One of the oldest approaches to integrating agents with other technologies is the use of \emph{wrappers} or \emph{transducers} that make the functionality of all the tools to be interconnected available through agent communication \cite{GeneserethKetchpel}. The overall system coordination can then be treated as a pure multi-agent system coordination problem. However, this approach has not gained traction in industry and we do not see it as a viable approach for integrating agents into enterprise computing environments.

A pragmatic but low-level approach for integrating agents with external systems is to call them directly from the agent program. If an agent platform is a framework for using a mainstream programming language for agent development (e.g. JADE\footnote{\url{http://jade.tilab.com/}}), then it is possible for agents to use whatever protocols and client libraries are supported in that language to invoke external services directly from within agents or to monitor for external events. An interpreter for a specialised agent programming language may allow user-defined code in the underlying implementation language to implement functionality called by the agent program. For example, new ``internal actions'' for Jason can be developed in Java, and these can use any Java communication libraries for external interaction. An agent's environment abstraction is another potential location for user customisation. For example, a Jason developer can implement an environment class that acts as a facade for external interaction.

The integration of agents with web services has been an important topic over the last decade, and some agent platforms provide specific support for this. For example, the online documentation for the JADE platform includes tutorials on calling web services from JADE and exposing agent services as web services, and the Jack WebBots \cite{JackWebBots} framework allows web applications to be built using agents.

More generally, it would be possible for the developers of an agent platform (or its community) to provide support for connecting agents to a range of external resource and service types. For example, the IMPACT agent platform \cite{IMPACT_JIIS_2000} includes a module that provides a uniform interface for connecting agents to external services, with support for a small number of service types already implemented.

The A\&A (Agents and Artifacts) meta-model extends the concept of an agent environment to include \emph{artifacts}. These represent resources and tools with observable properties and specific operations that agents can invoke. These can be used to provide services internal to an MAS, or as an interface to external services, such as web services \cite{JAAMAS_artifacts}. However, it is unlikely that the developer and user community for any agent-specific technology, whether a specific platform like IMPACT or a more general approach such as A\&A, could rival the scale and diversity provided by a more mainstream integration technology such as Camel, which supports more than 130 endpoint types. Also, for the case of A\&A, an agent developer would need to learn multiple APIs (for each artifact type) when integrating agents with different types of external service. This is not the case in our approach (see Section~\ref{sec:conclusion}).

The \emph{active components} paradigm is a combination of a component model with agent concepts \cite{ActiveComponentsMATES10}. Active components can communicate via method calls or asynchronous messages and may be hierarchically composed of subcomponents. They run within a management infrastructure that controls non-functional properties such as persistence and replication. They may have internal architectures of different types, and this heterogeneity, combined with a uniform external interface model, facilitates the interoperation of different types of system that are encapsulated as active components. As with artifacts, the success of this approach for large-scale integration rests on the availability of active components encapsulating a wide range of service types. However, a Camel context could be encapsulated within an active component (or, alternatively, an artifact).

\section{Conclusion}
\label{sec:conclusion}

In this paper we have proposed a novel approach for integrating agents with external resources and services by leveraging the capabilities of existing enterprise integration technology. By using a mainstream technology we can benefit from the competitive market for robust integration tools (or the larger user base for open source software), and can have access to a much larger range of pre-built components for connecting to different resource and service types. This is evidenced by Camel's large number of available endpoint types.

We presented the design of an interface between agents and the Camel integration framework in terms of the EIP endpoint concept. This can serve as a pattern for interconnecting agents with any type of message-based middleware.

We described an implemented architecture for this approach and illustrated its practical use in a hypothetical (but, we think, plausible) business process use case. The Camel routes we presented demonstrate the benefits of using a specialist coordination tool such as Camel for handling the coordination of distributed agents and services, and leaving the agent code to provide the required core functionality. This division of responsibilities also enables a division of implementation effort: the coordination logic can be developed by business process architects using a programming paradigm that directly supports common enterprise integration patterns, and less development time is needed from (currently scarce) agent programmers. An agent programmer using our framework does not need to learn any APIs for client libraries or protocols---the agent code can be based entirely on the traditional agent concepts of messages, actions and plans. The developer of the message-routing logic does not need to know much about agents except the basic concepts encoded in the agent endpoint design (\emph{message}, \emph{illocutionary force}, \emph{action}, \emph{percept}, etc.) and the syntax of the agent messages to be sent from and received by the message routes.

\bibliographystyle{splncs03}
\bibliography{aamas2013}

\begin{thebibliography}{1}
\providecommand{\url}[1]{\texttt{#1}}
\providecommand{\urlprefix}{URL }

\bibitem{JackWebBots}
{Agent Oriented Software}: {JACK Intelligent Agents} {WebBot} manual.
  \url{http://www.aosgrp.com/documentation/jack/WebBot_Manual_WEB/} (2011)

\bibitem{JasonBook}
Bordini, R.H., {H\"ubner}, J.F., Wooldridge, M.: Programming Multi-Agent
  Systems in {AgentSpeak} using {Jason}. Wiley (2007)

\bibitem{ESBbook}
Chappell, D.: {Enterprise Service Bus}: Theory in Practice. O'Reilly (2004)

\bibitem{GeneserethKetchpel}
Genesereth, M.R., Ketchpel, S.P.: Software agents. Communications of the ACM
  37(7),  48--53 (1994)

\bibitem{HohpeWoolf2004}
Hohpe, G., Woolf, B.: Enterprise Integration Patterns: Designing, Building, and
  Deploying Messaging Solutions. Addison-Wesley (2004)

\bibitem{CamelBook}
Ibsen, C., Anstey, J.: Camel in Action. Manning (2010)

\bibitem{ActiveComponentsMATES10}
Pokahr, A., Braubach, L., Jander, K.: Unifying agent and component concepts.
  In: Multiagent System Technologies, LNAI, vol. 6251, pp. 100--112. Springer
  (2010)

\bibitem{JAAMAS_artifacts}
Ricci, A., Piunti, M., Viroli, M.: Environment programming in multi-agent
  systems: an artifact-based perspective. Autonomous Agents and Multi-Agent
  Systems  23(2),  158--192 (2011)

\bibitem{IMPACT_JIIS_2000}
Rogers, T.J., Ross, R., Subrahmanian, V.: {IMPACT}: A system for building agent
  applications. Journal of Intelligent Information Systems  14,  95--113 (2000)

\end{thebibliography}

\end{document}